\documentstyle[aas2pp4]{article}

\def\ltsima{$\; \buildrel < \over \sim \;$}
\def\simlt{\lower.5ex\hbox{\ltsima}}            
\def\gtsima{$\; \buildrel > \over \sim \;$}
\def\simgt{\lower.5ex\hbox{\gtsima}}            

\newcommand{\ci}{CI~Cam}
\newcommand{\xte}{XTE J0421+560}
\newcommand{\rxte}{{\it RXTE}}
\newcommand{\asca}{{\it ASCA}}
\newcommand{\etal}{{\it et al.}}

\newcommand{\erg}{erg s$^{-1}$}

\slugcomment{Accepted for publication in APJL}
\lefthead{Ueda et al.}
\righthead{ASCA Observation of \ci}

\begin{document}

\title{ASCA Observation of the Galactic Jet Source \xte\ (\ci) in Outburst}
\author{Y. Ueda\altaffilmark{1}, 
M. Ishida\altaffilmark{1},
H. Inoue\altaffilmark{1}, 
T. Dotani\altaffilmark{1},
J. Greiner\altaffilmark{2},
W.H.G. Lewin\altaffilmark{3}
}

\altaffiltext{1}{Institute of Space and Astronautical Science, Yoshinodai 3-1-1, Sagamihara, Kanagawa 229-8510, Japan}
\altaffiltext{2}{Astrophysikalisches Institut Potsdam, An der Sternwarte 16, 14482 Potsdam, Germany}
\altaffiltext{3}{Department of Physics, Center for Space Research, Massachusetts Institute of Technology, Cambridge, MA 02139, U.S.A}

\begin{abstract}

We observed the newly discovered Galactic jet source \xte\ (= \ci)
with \asca\ from 1998 April 3.3 to April 4.1 (UT), three days after
the beginning of the outburst. The X-ray intensity in the 1--10 keV
band gradually decreased with an e-folding time of about 30 hours; the
decline was accompanied by spectral softening. Two flare-like
intensity enhancements were detected below $\sim$1 keV. We could fit
the average spectrum above 0.8 keV with a two-temperature model (5.7
and 1.1 keV) of thermal emission from an optically thin
ionization-equilibrium plasma. The broad iron-K profile, however,
requires an extra emission line at 6.4 keV, or Doppler broadening (or
both). The former can be explained in terms of reflection from cold
matter, while the latter can be attributed to emission from the twin
jets. In both cases, the time evolution of the emission measure and of
the temperature are difficult to explain by emission from a single
plasma, suggesting that heat input and/or injection of material was
occurring during the outburst.

\end{abstract}


\twocolumn

\section{Introduction}

\xte\ is an X-ray transient discovered by the All Sky Monitor (ASM) on board
the {\it Rossi} X-ray Timing Explorer (\rxte) on 1998 March 31.6 (Smith
\etal\ 1998). After the detection, its X-ray intensity (2--12 keV)
rapidly rose and reached a peak of about 2 Crab on 1998 April
1.04-1.08 (UT), and decayed with an e-folding time of 0.6 day (Belloni
\etal\ 1998). The hard X-rays in the 20--70 keV band were also detected with
BATSE on board {\it CGRO}, accompanied with a similar intensity
evolution (Harmon, Fishman, \& Paciesas 1998).

Subsequent radio and optical observations identified \xte\ with the
previously known symbiotic star \ci (= MWC~84, see e.g., Downes 1984). 
On April 1.9 and 2.6, a new variable radio source was detected in the
ASM X-ray error box with the Very Large Aray (VLA), at a position that
coincides with \ci\ (Hjellming \& Mioduszewski 1998a; 1998b). Optical
observations revealed brightening of \ci\ by 3.5 magnitude in the
R-band (Robinson \etal\ 1998), and the presence of a He II emission
line in its spectra (Wagner \& Starrfield 1998; Garcia \etal\ 1998).
More interestingly, VLA images taken on April 5.1 and 6.9 showed
nearly symmetric twin jets (Hjellming and Mioduszewski 1998c) that
were similar to those of SS~433 (Margon 1984), expanding with an
apparent velocity of 0.15$c$ ($c$ is the light speed) for an assumed
distance of $d=$1 kpc (see Bergner \etal\ 1995).

The outburst of this new Galactic jet source, characterized by a rapid
rise and decay, is unusual among other X-ray transients (Chen, Shrader, \& Livio 1997). In this paper, we report results from \asca\ observations of
\xte, performed about 3 days after the onset of the outburst. We
discuss the origin of the X-ray emission and implications for its
relation to the jets.

\section{Observation and Analysis}

Following the discovery by \rxte, we observed \xte\ with the \asca\
satellite (Tanaka, Inoue, \& Holt 1994) from 1998 April 3.31 (= MJD
50906.31) to April 4.14 as a Target of Opportunity (TOO) observation
(Ueda \etal\ 1998). The GIS (Ohashi \etal\ 1996) was operated in the
pulse-height mode and the SIS (Burke \etal\ 1994) was in the bright mode.
A net exposure of 39,000 sec was achieved after standard data
selection. The time-averaged GIS count rate corrected for dead time
was 20 c s$^{-1}$ per sensor ($r<6'$).

\begin{figure}[hpt]
\plotone{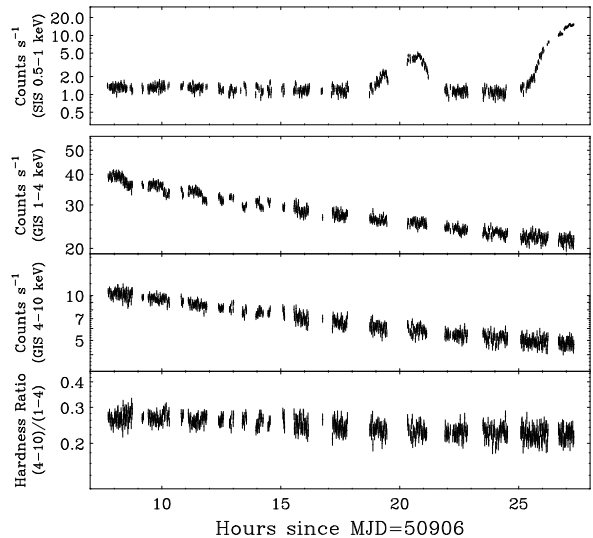}
\caption{
Light curves of \xte\ obtained with \asca\ in three energy bands,
0.5--1 keV (1st panel), 1--4 keV (2nd panel), and 4--10 keV (3rd
panel), and the spectral hardness ratio between the 1--4 keV and 4--10
keV bands (4th panel), binned at 64 s. The count rate of SIS1 within a
selected region is plotted for the 0.5--1.0 keV band, while the sum of
the GIS2 and GIS3 count rates, within a radius of 6$'$, are shown for
the other bands. Dead-time corrections were performed for the GIS
data.
\label{LightCurve}}
\end{figure}

Figure~1 shows the light curve in three energy bands, 0.5--1 keV, 1--4
keV, and 4--10 keV, together with the spectral hardness ratio between
the 1--4 keV and 4--10 keV bands. Here, we used the GIS data for the
1--4 and 4--10 keV bands, but used the SIS data for the 0.5--1 keV
band, considering its superior detection efficiencies in that band. 
The intensities above 1 keV gradually decreased, accompanied by a
slight spectral softening. Using the GIS count rate in the 1--10 keV
range, we obtained an e-folding decay time of 30.4$\pm$0.3 hours.  On
the other hand, two flare-like enhancements are noticed in the 0.5--1
keV band, which began around April 3 19:00 and April 4 1:00. The
behavior of these flares seem to be completely independent of the
light curves above 1 keV, suggesting a different origin from the
steadily decaying emission.
\placefigure{LightCurve}

\begin{figure}[hpt]
\plotone{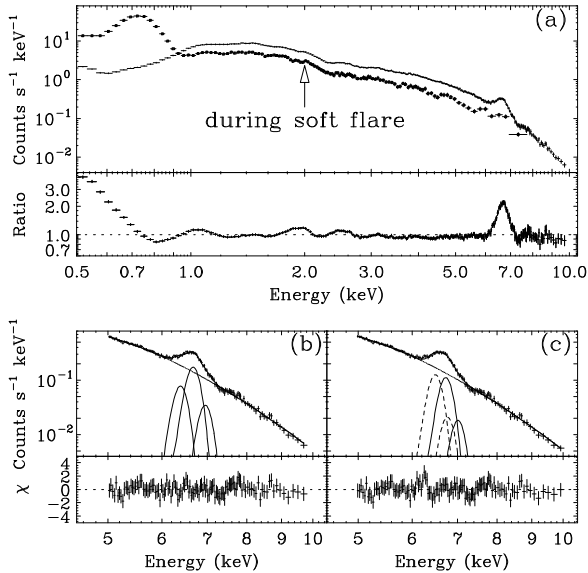}
\caption{
(a): 
The SIS folded spectrum of \xte\ (SIS0+SIS1) averaged from 1998 April 3
7:30 to April 4 2:40 (no symbol).  Lower panel shows the ratio of the
data to the best-fit thermal bremsstrahlung model determined in the
0.8--10 keV band. The spectrum taken from April 4 2:40 to 3:20 is
plotted with square symbols. 
%\newline
(b): 
The fitting results of the SIS spectrum in the 5--10 keV band with a
thermal bremsstrahlung spectrum plus 3 narrow lines. The model of each
component is plotted separately. We fixed the line energies at 6.4
keV, 6.68 keV, and 6.96 keV. The residuals are plotted in the lower
panel.
%\newline
(c): 
Same as (b) but with 4 narrow lines, consisting of 2 red-shifted lines
(dashed curves) and 2 blue-shifted lines with rest energies of 6.68
keV and 6.96 keV. The line-intensity ratio between the 6.96 keV and
the 6.68 keV lines are fixed at 0.2, which is expected from a plasma
with a temperature of about 6 keV.
\label{energySpetrum}
}
\end{figure}

We first examined spectra averaged over the whole observation
including the times of the soft flares (except from April 4 2:40 to
3:20 for the SIS, when the telemetry saturation is significant.)  We
neglected the background, which is negligible (0.02\%) compared with
the source flux. We limited the energy range of the GIS spectrum used
in the spectral fitting to 3.2--10 keV, in order to avoid possible
uncertainties in the low energy response (Ishisaki \etal\ 1998). 
\placefigure{energySpetrum}

Figure~2 (a) shows the time-averaged raw (folded) spectrum of the SIS
(SIS0+SIS1). In the lower panel we show the ratio of the data fit to a
simple continuum model in the 0.8--10 keV range. For
the continuum, we adopted a thermal bremsstrahlung spectrum with
interstellar absorption. We can see a soft-excess below 0.8 keV, which
is attributable to the soft-flares. Above 0.8 keV, on the other hand,
a strong emission-line feature is clearly seen around 6.7 keV, which
probably corresponds to iron K lines. Also, the humps around 1.2 keV,
1.9 keV and 2.5 keV are suggestive of the iron L-line forest, K-lines
from silicon (1.86 keV for Si~XIII and 2.01 keV for Si~XIV) and sulfur
(2.45 keV for S~XIV and 2.62 keV for S~XV), respectively.

As the spectral features discussed above are probably of an
optically-thin thermal plasma origin, we attempted to fit the SIS
spectrum in the 0.8--10 keV range and the GIS spectrum in the 3.2--10
keV range simultaneously with an ionization-equilibrium plasma
emission model given by Raymond \& Smith (1977, hereafter RS model),
corrected for interstellar absorption. We set elemental abundances
of Si, S, and other elements as three free parameters, while that of
He is fixed at the Solar value. A single temperature RS model,
however, could not reproduce the data ($\chi^2/\nu$ = 1578/762 for an
obtained temperature of 5.4 keV). Significant structures remained
around 2 keV, 2.5 keV, and 6.4 keV in the residuals.

The poor fit in the Si and S K-bands arises because their observed
line-intensity ratio between helium-like and hydrogen-like ions can
not be produced by a plasma with a temperature of $\sim$5 keV. More
directly, we determined these ratios by fitting the spectra with a
model consisting of the lines from H-like and He-like ions (modeled by
narrow Gaussians) and a continuum (modeled by a thermal
bremsstrahlung spectrum). The obtained ratios, 0.69$\pm$0.12 for (Si~XIV /
Si~XIII) and 0.54$\pm$0.18 for (S~XIV / S~XV), indicate corresponding
ionization-equilibrium temperatures of 1.2$\pm$0.2 keV and 1.6$\pm$0.3
keV, respectively (Mewe, Gronenschild,
\& van den Oord 1985).  This suggests that the observed emission originates
from a multi-temperature plasma. It is supported by a great improvement
in the fit with a two-temperature RS model (hereafter 2T-RS model),
which gives $\chi^2/\nu$ = 1232/760 for obtained temperatures of
5.7$\pm$0.1 keV and 1.06$\pm$0.02 keV.

The 2T-RS model, however, still could not reproduce the line profile
in the iron K-band, leaving large positive residuals at $\sim$ 6.4
keV.  This suggests (I) the presence of an unresolved line around 6.4
keV, and/or (II) Doppler broadening, which could be attributed to the
jets. Accordingly, we fit the SIS spectrum in the 5--10 keV band with
two simplified models: (I) a thermal bremsstrahlung spectrum plus 3 narrow
lines at 6.4 keV, 6.68 keV (K$\alpha$ from Fe~XXV), and 6.96 keV
(K$\alpha$ from Fe~XXVI), and (II) a thermal bremsstrahlung spectrum plus 4
narrow lines, consisting of 2 red-shifted lines and 2 blue-shifted
lines with rest energies of 6.68 keV and 6.96 keV.  Both
models can reproduce the iron line profile (Figure~2(b) and 2(c)).  We
applied the following models to the whole spectra: (Model~I) 2T-RS
model with a narrow emission line around 6.4 keV, and (Model~II) 2T-RS
model consisting of two components with common spectral parameters
(with the same normalizations, assuming that Doppler boosting is
negligible), each modified by a red-shift and a blue-shift. We found
that both models gave acceptable fits ($\chi^2/\nu$ = 619/759 for
Model~I and 658/759 for Model~II). Two temperatures are still
necessary for Model~II in order to explain the line profiles of Si and
S. The best fit parameters are summarized in Table~1. From Model~I, we
obtained an energy of 6.41$\pm$0.04$\pm$0.06 keV and an equivalent
width of 90$\pm$11$\pm$30 eV for the narrow line (the second error
represents the systematic error due to the uncertainty in the absolute
gain (1\%, Makishima
\etal\ 1996)). The line center energy is consistent with that of
K$\alpha$ from cold iron. Model~II gave 
redshift parameters (defined
as $\delta
\lambda/\lambda$) of $z_+ = 0.029 (\pm0.002\pm0.010)$ for the red-shifted
component, and $z_- = -0.007 (\pm0.002\pm0.010)$ for the blue-shifted
component. 
\placetable{tbl-1}

Since the spectrum evolves in time (see Figure~1), we divided the
observation into four epochs (April 3 7:00 -- 12:00, 12:00 -- 18:00,
18:00 -- 23:00, and April 3 23:00 -- April 4 3:20) and examined their
spectra above 0.8 keV separately. We fit them with Model~I, fixing the
elemental abundances at the best-fit values obtained from the averaged
spectra (Table~1). Figure~3 shows the time history of the temperatures and
emission measures separately for the high and low temperature
components. We also calculated the summed emission measure from the
two components and their mean temperature ($T$) weighted with the
emission measure, as representative parameters for the plasma in each
epoch. The results are plotted in Figure~3.

Figure~3 clearly indicates that both the summed emission measure,
$E$, and the mean temperature, $T$, decreased with
time. To be quantitative, we fit these curves in a form of $E(t)
\propto t^{-\alpha}$ and $T(t) \propto t^{-\beta}$, where $t$ is
defined as a time in units of hours since the beginning of the
outburst ($t_0$).  Assuming $t_0$ to be March 31.6 (the onset of the
X-ray outburst, Belloni \etal\ 1998), we obtained $\alpha$=2.39$\pm$0.23
and $\beta$=0.95$\pm$0.16. The best-fit curves are plotted with dashed
lines in Figure~3. When we take $t_0$ to be April 1.04 (the peak of
the X-ray outburst), the indices become $\alpha$=2.05$\pm$0.20 and
$\beta$=0.82$\pm$0.14, respectively.
\placefigure{SpecralEvolution}
\begin{figure}[hpt]
\plotone{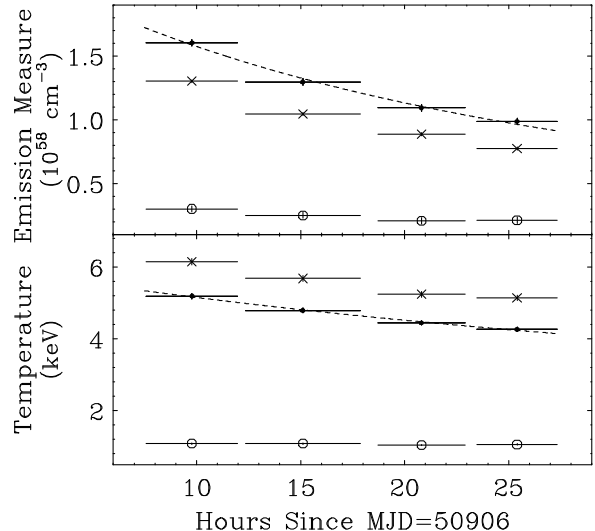}
\caption{
History of the emission measures (in units of $10^{58} (d/{\rm 1
kpc})^2$ cm$^{-3}$) and the temperatures, derived from the 2T-RS
model.  Crosses and open circles represent those for the high- and the
low-temperature components, respectively. The sum of two emission
measures and the averaged temperature weighted with the emission
measure are plotted with small filled circles. The dashed lines
represent the best-fit models in the form of a power law as a function
of time since the onset of the outburst (March 31.6).
\label{SpectralEvolution}
}
\end{figure}

Finally, we made the SIS spectrum in the epoch from April 4 2:40 to
3:20, when the soft-flare was the most prominent, in order to study
the origin of the soft flares (Figure~2(a)).
The soft-excess component over the extrapolation from higher energies
is found to have a sharp cut-off around 0.8 keV.
We could fit it with an absorbed blackbody model modified with two
deep absorption edges, although it is hard to reject other
possibilities (e.g., the superposition of multiple lines) due to the
limited energy resolution. We obtained a temperature of 0.12$\pm$0.02
keV, an unabsorbed bolometric luminosity of
$(1.7^{+2.0}_{-0.4})\times10^{37}$ $(d/{\rm 1 kpc})^2$ \erg, the
hydrogen column density (in excess of the interstellar absorption) of
$N_{\rm H}$ = $(2.1^{+1.2}_{-0.4}) \times 10^{21}$ cm$^{-2}$, and edge
energies of 0.77$\pm$0.02 keV and 0.84$\pm$0.03 keV with optical
depths of 2.3$^{+1.1}_{-0.7}$ and 26$\pm$6, respectively. The radius
of the emitting region is then estimated at
$(1.4^{+0.8}_{-0.2})\times10^{8} (d/{\rm 1 kpc})$ cm.

\section{Discussion}

We have shown that the overall spectrum of \xte\ in the
0.8--10 keV range can be represented by the emission from an
optically-thin thermal plasma with at least two temperatures. A
modification is necessary, however, to account for the broad iron-K
line profile. We thus propose two models (Model~I and II), which
cannot be distinguished from the data.

If Model~I is correct, we need to explain the large ($\sim$90 eV)
equivalent width (E.W.) of the 6.4 keV line. A fluoresence iron line
with an E.W. of $\sim$100 eV can be produced if cold matter
spherically surrounds the X-ray emitter with a column density of
$N_{\rm H} \sim 10^{23}$ cm$^{-2}$ (e.g., Inoue 1985). This situation
is unlikely, because the observed column density is at most
$5\times10^{21}$ cm$^{-2}$, unless we were seeing less than 5\% of the
direct component. Another possibility is that we see the reflection from
optically-thick cold matter. The observed E.W. can then be explained if the
solid angle of the reflector seen from the emitter, $\Omega$, is
$\sim2\pi$ (e.g., Basko 1978). In this case, we should see an
accompanying reflection component, which has an absorption edge
structure above 7.1 keV (e.g., Lightman \& White 1988; George \&
Fabian 1991). Applying the standard edge model with Model~I, we found
that the presence of an absorption edge at 7.1$\pm$0.2 keV is allowed
with an optical depth of 0.06$\pm$0.02, which roughly
corresponds to $\Omega =2\pi\sim4\pi$ (Lightman \& White 1988).

Based on Model~I, we first examine the situation that a static plasma
was cooling down by radiation. In this case, the decay time-scale of
the emission, $\tau$, is given by that of the radiative cooling, $kT /
n \Lambda (T)$, where $T$ is the temperature, $n$ is the number
density (assumed to be common for electrons and ions), and $\Lambda
(T)$ is the emissivity of the plasma. Using $\tau \sim 30$ hours and
$kT\sim 10$ keV, we obtain $n \sim 10^{10}$ cm$^{-3}$. Combined with
the observed emission measure ($n^2 V\sim10^{58}$ cm$^{-3}$), the size
of the plasma, $r\sim V^{1/3}$, is estimated at $\sim 10^{13}$ cm. 
This seems to be too large, however, to find the origin of the
reflector in the system of a compact object (e.g., surface of a white
dwarf), which must cover the plasma with $\Omega\sim2\pi$. Thus, to
make this picture consistent, input of heat and/or material into the
plasma must be considered.

If Model~II applies (this means \xte\ is a second example of a
Galactic jet source where we directly see the X-ray emission from the
jets, as is the case for SS~433), we can constrain the geometry of the
jets, using the redshift parameters derived from the spectral fitting.
The redshift parameters, $z_+$ (red-shifted component) and $z_-$
(blue-shifted component), are expressed as $z_{\pm}
(\equiv\frac{\Delta \lambda_{\pm}}{\lambda}) = \gamma (1 \pm
\frac{v}{c}~{\rm cos}~\theta) - 1 $, where $v$ is the velocity of jets
(assumed to be the same for the twin jets), $\gamma \equiv
[1-(v/c)^2]^{1/2}$ is the Lorentz factor, and $\theta$ is the angle
between the line of sight and the axis of the jet.  Their mean value,
$(z_+ + z_-)/2 = \gamma-1$, gives the traverse Doppler effect (time
dilation). Our result in Table~1 yields $(z_+ + z_-)/2 =
(1.1\pm0.2\pm1.0)\times10^{-2}$, which corresponds to $v =
0.15^{+0.06}_{-0.11} c$. The quantity $(z_+ - z_-)/2 =
\gamma$~$(v/c)$~cos~$\theta$ can be used to constrain $\theta$. The
observed value, $(1.8\pm0.2\pm0)\times10^{-2}$, indicates cos~$\theta
= 0.12\pm 0.02$, hence $\theta=(83\pm2)^\circ$, assuming $v=0.15 c$.

The energy conservation equation for an expanding plasma which is
uniform and isothermal is represented by:
$$
        \frac{dU}{dt} = - \Lambda (T)\; n^2 V - n kT \frac{dV}{dt},
$$
where $k$ is the Boltzmann constant, $V$ is the volume, $U = (3/2) n
kT V$ is the total energy in the plasma (here we have assumed a
constant velocity), and $\Lambda (T)$ is the emissivity of the plasma. 
The first term on the right represents the radiative cooling, while
the second represents the adiabatic cooling.  First, we assume that
the total number of electrons in the plasma, $N (\equiv n V)$, is
conserved. To compare this with the previous analysis, we represent the
time dependence of the emission measure and the temperature in a form
of $E (t) \propto t^{-\alpha}$ and $ T (t)\propto t^{-\beta}$, where
$t$ is the time since the expansion started. This yields $n \propto
t^{-\alpha}$, and $V \propto n^{-1}
\propto t^{\alpha}$. Note that $\alpha = 3$ corresponds to the
3-dimensional self-similar free expansion. Finally, the energy conservation
equation becomes:
$$
	-\frac{3}{2} \beta \frac{kT}{t} = -n\Lambda(T) - \alpha \frac{kT}{t}.  $$
We found, however, that this equation cannot be satisfied for the observed
values of $\alpha$ = 2.39$\pm0.23$ and $\beta$ = 0.95$\pm$0.16,
because the left term is always smaller than the second term on the
right.
This implies that the plasma we observed is not a single, freely
expanding plasma. Injection of new (hot) material, and/or heat input,
such as irradiation from the compact star, are required to prevent the
observed temperature from cooling fast. An example of the former 
can be found in the cooling jet model applied to SS~433 (Kotani 1997).

The origin of the soft flares seems to be independent of that of the
thin thermal emission discussed above. The spectrum of the soft flares
are similar to that observed from the super soft sources (SSSs, see
e.g., Greiner 1996). The apparent edges might be attributable to those
of O~VII (0.74 keV) and O~VIII (0.87 keV) (Ebisawa \etal\ 1996). 
However, a temperature as high as 120 eV and the rapid time
variability on a time scale of 1 hour have not been observed from the
SSSs. We leave it for future studies to find out what the soft flares
are and what causes the soft flares.

In summary, we revealed that the X-ray emission from \xte\ came from a
multi-temperature thin-thermal plasma. The iron line profile suggests
the presence of a reflection component and/or the interesting
possibility that this emission is produced in the twin jets, as is the
case for SS~433. It will be of great interest to examine this further
with radio data.

\acknowledgments

We thank members of the ASCA team for making this TOO observation
possible, and Y.~Tanaka for useful discussions. JG is supported by the
German Bundesmi\-ni\-sterium f\"ur Bildung, Wissenschaft, Forschung
und Technologie (BMBF/DLR) under contract No.FKZ 50 QQ 9602 3. WHGL is
grateful to NASA for support.

%\clearpage
\begin{deluxetable}{p{5cm}lll}
\small
\tablenum{1}
\tablecaption{Results of the Spectral Fits\label{tbl-1}}
\tablewidth{0pt}
\tablehead{
\colhead{}&\colhead{Model~I\tablenotemark{a}}&\colhead{Model~II\tablenotemark{b}}}

\startdata
$N_{\rm H}$ ($10^{21}$ cm$^{-2}$)& 4.6$\pm$0.3 & 4.3$\pm$0.3\nl
T$_1$ (keV)& 5.68$\pm$0.11 & 5.65$\pm$0.08 \nl
EM$_1$\tablenotemark{c}& 0.83$\pm$0.02 & 0.83$\pm$0.02\nl
T$_2$ (keV)& 1.07$\pm$0.03 & 1.10$\pm$0.04 \nl
EM$_2$\tablenotemark{c}& 0.21$\pm$0.03 & 0.14$\pm$0.03\nl
Si Abundance\tablenotemark{d}& 1.25$\pm$0.15 & 1.67$\pm$0.19\nl
S Abundance\tablenotemark{d}& 1.03$\pm$0.13 &  1.29$\pm$0.16\nl
Others Abundance\tablenotemark{d}& 0.36$\pm$0.02 & 0.46$\pm$0.02\nl
Line Energy (keV)& 6.41 $\pm$0.04$\pm$0.06\tablenotemark{e} &\nodata\nl
Equivalent Width (eV)& 90$\pm$11$\pm$30\tablenotemark{e} &\nodata\nl
$z_+$ \tablenotemark{f} & \nodata & 2.9$\pm$0.2$\pm$1.0\tablenotemark{e}\nl
$z_-$ \tablenotemark{f} & \nodata & -0.68$\pm$0.18$\pm$1.0\tablenotemark{e}\nl
$\chi^2/\nu$& 619/759& 658/759\nl

\tablenotetext{a}{Two-temperature Raymond \& Smith plasma model with an emission line.}
\tablenotetext{b}{Two-temperature Raymond \& Smith plasma model consisting of
twin Doppler-shifted components with common parameters.}
\tablenotetext{c}{Emission measure in units of $10^{58}$ $(d/{\rm 1 kpc})^2$ cm$^{-3}$ (sum of the twin components for Model~II). }
\tablenotetext{d}{Relative to the Solar abundance as given by Anders \& Grevesse (1989). }
\tablenotetext{e}{The first error represents the statistical error; the second one represents the systematic error due to the uncertainty in the gain.}
\tablenotetext{f}{Redshift parameter ($\equiv \Delta \lambda/\lambda$) in units of $10^{-2}$. $z_+$ and $z_-$ correspond to the red-shifted and the blue-shifted components, respectively.}
\tablecomments{Errors are 90\% confidence limits for a single parameter.}
\enddata
\end{deluxetable}

\end{document}